\documentclass[a4paper,fleqn,usenatbib]{mnras}

\usepackage[T1]{fontenc}
\usepackage{ae,aecompl}
\usepackage{hyperref}

\usepackage{graphicx}	
\usepackage{amsmath}	
\usepackage{amssymb}	
\usepackage{multirow}

\newcommand{\hsmsun}{\,h_{70}^{-2} {\rm M_\odot}}
\newcommand{\hskpc}{\, h_{70}^{-1}{\rm{kpc}} }
\newcommand{\hsMpc}{\, h_{70}^{-1}{\rm{Mpc}} }

\newcommand{\lan}{\langle}
\newcommand{\ra}{\rangle}

\newcommand{\lcdm}{{\rm \Lambda CDM}}

\newcommand*{\swap}[2]{#2#1}

\title[Lensing test of Verlinde's Emergent Gravity]{First test of Verlinde's theory of Emergent Gravity using Weak Gravitational Lensing measurements}
\author[M. M. Brouwer et al.]{Margot M. Brouwer$^{1}$\thanks{E-mail:brouwer@strw.leidenuniv.nl}, 
	Manus R. Visser$^{2}$,
	Andrej Dvornik$^{1}$, 
	Henk Hoekstra$^{1}$, \and
	Konrad Kuijken$^{1}$,
	Edwin A. Valentijn$^{3}$,
	Maciej Bilicki$^{1}$,
	Chris Blake$^{4}$, \and
	Sarah Brough$^{5}$,
	Hugo Buddelmeijer$^{1}$, 
	Thomas Erben$^{6}$, 
	Catherine Heymans$^{7}$, \and
	Hendrik Hildebrandt$^{6}$,
	Benne W. Holwerda$^{1}$,
	Andrew M. Hopkins$^{5}$, 
	Dominik Klaes$^{6}$, \and
	Jochen Liske$^{8}$,
	Jon Loveday$^{9}$,
	John McFarland$^{3}$,
	Reiko Nakajima$^{6}$, \and
	Crist\'obal Sif\'on$^{1}$,
	Edward N. Taylor$^{4}$.
	\\
	\\
	$^{1}$Leiden Observatory, Leiden University, Niels Bohrweg 2, 2333 CA Leiden, The Netherlands.\\
	$^{2}$Institute for Theoretical Physics, University of Amsterdam, Science Park 904, 1098 XH Amsterdam, The Netherlands. \\
	$^{3}$Kapteyn Astronomical Institute, University of Groningen, P.O. Box 800, 9700 AV Groningen, The Netherlands.\\
	$^{4}$Centre for Astrophysics and Supercomputing, Swinburne University of Technology, Hawthorn 3122, Australia. \\
	$^{5}$Australian Astronomical Observatory, P.O. Box 915, North Ryde, NSW, Australia.\\
	$^{6}$Argelander-Institut f{\"u}r Astronomie, Auf dem H{\"u}gel 71, D-53121 Bonn, Germany.\\
	$^{7}$SUPA, Institute for Astronomy, University of Edinburgh, Royal Observatory, Blackford Hill, Edinburgh, EH9 3HJ, UK.\\
	$^{8}$Hamburger Sternwarte, Universit\"at Hamburg, Gojenbergsweg 112, 21029 Hamburg, Germany. \\
	$^{9}$Astronomy Centre, University of Sussex, Falmer, Brighton BN1 9QH, UK. \\
}

\date{Accepted XXX. Received YYY; in original form ZZZ}

\pubyear{2016}

\begin{document}
\label{firstpage}
\pagerange{\pageref{firstpage}--\pageref{lastpage}}
\maketitle

\begin{abstract}
Verlinde (2016) proposed that the observed excess gravity in galaxies and clusters is the consequence of Emergent Gravity (EG). In this theory the standard gravitational laws are modified on galactic and larger scales due to the displacement of dark energy by baryonic matter. EG gives an estimate of the excess gravity (described as an \emph{apparent} dark matter density) in terms of the baryonic mass distribution and the Hubble parameter. In this work we present the first test of EG using weak gravitational lensing, within the regime of validity of the current model. Although there is no direct description of lensing and cosmology in EG yet, we can make a reasonable estimate of the expected lensing signal of low redshift galaxies by assuming a background $\lcdm$ cosmology. We measure the (apparent) average surface mass density profiles of $33,613$ isolated central galaxies, and compare them to those predicted by EG based on the galaxies' baryonic masses. To this end we employ the $\sim 180$ deg$^2$ overlap of the Kilo-Degree Survey (KiDS) with the spectroscopic Galaxy And Mass Assembly (GAMA) survey. We find that the prediction from EG, despite requiring \emph{no free parameters}, is in good agreement with the observed galaxy-galaxy lensing profiles in four different stellar mass bins. Although this performance is remarkable, this study is only a first step. Further advancements on both the theoretical framework and observational tests of EG are needed before it can be considered a fully developed and solidly tested theory.
\end{abstract}


\newpage
\clearpage



\section{Introduction}
\label{sec:introduction}
In the past decades, astrophysicists have repeatedly found evidence that gravity on galactic and larger scales is in excess of the gravitational potential that can be explained by visible baryonic matter within the framework of General Relativity (GR). The first evidence through the measurements of the dynamics of galaxies in clusters {\cite[]{zwicky1937}} and the Local Group {\cite[]{kahn1959}}, and through observations of galactic rotation curves (inside the optical disks by {\citealp{rubin1983}}, and far beyond the disks in hydrogen profiles by \citealp{bosma1981}) has been confirmed by more recent dynamical observations \cite[]{martinsson2013,rines2013}. Furthermore, entirely different methods like gravitational lensing {\cite[]{hoekstra2004,mandelbaum2015,linden2014,hoekstra2015}} of galaxies and clusters, Baryon Acoustic Oscillations {\cite[BAO's,][]{eisenstein2005, blake2011}} and the cosmic microwave background {\cite[CMB,][]{spergel2003,planck2015}} have all acknowledged the necessity of an additional mass component to explain the excess gravity. This interpretation gave rise to the idea of an invisible \emph{dark matter} (DM) component, which now forms an important part of our standard model of cosmology. In our current $\lcdm$ model the additional mass density (the density parameter $\Omega_{\rm CDM} = 0.266$ found by {\citealp{planck2015}}) consists of cold (non-relativistic) DM particles, while the energy density in the cosmological constant ($\Omega_{\rm \Lambda} = 0.685$) explains the observed accelerating expansion of the universe. In this paradigm, the spatial structure of the sub-dominant baryonic component (with $\Omega_{\rm b} = 0.049$) broadly follows that of the DM. When a DM halo forms through the gravitational collapse of a small density perturbation {\cite[]{peebles1970structureform}} baryonic matter is pulled into the resulting potential well, where it cools to form a galaxy in the centre {\cite[]{white1978galaxyform}}. In this framework the excess mass around galaxies and clusters, which is measured through dynamics and lensing, has hitherto been interpreted as caused by this DM halo.

In this paper we test the predictions of a different hypothesis concerning the origin of the excess gravitational force: the \cite{verlinde2016} model of Emergent Gravity (EG). Generally, EG refers to the idea that spacetime and gravity are macroscopic notions that arise from an underlying microscopic description in which these notions have no meaning. Earlier work on the emergence of gravity has indicated that an area law for gravitational entropy is essential to derive Einstein's laws of gravity \cite[]{jacobson1995,padmanabhan2010,verlinde2011,faulkner2015,jacobson2016}. But due to the presence of positive dark energy in our universe \cite{verlinde2016} argues that, in addition to the area law, there exists a volume law contribution to the entropy. This new volume law is thought to lead to modifications of the emergent laws of gravity at scales set by the `Hubble acceleration scale' $a_0 = c H_0$, where $c$ is the speed of light and $H_0$ the Hubble constant. In particular, \cite{verlinde2016} claims that the gravitational force emerging in the EG framework exceeds that of GR on galactic and larger scales, similar to the MOND phenomenology \cite[Modified Newtonian Dynamics,][]{milgrom1983} that provides a successful description of galactic rotation curves \cite[e.g.][]{mcgaugh2016}. This excess gravity can be modelled as a mass distribution of \emph{apparent} DM, which is only determined by the baryonic mass distribution $M_b(r)$ (as a function of the spherical radius $r$) and the Hubble constant $H_0$. In a realistic cosmology, the Hubble parameter $H(z)$ is expected to evolve with redshift $z$. But because EG is only developed for present-day de Sitter space, any predictions on cosmological evolution are beyond the scope of the current theory. The approximation used by \cite{verlinde2016} is that our universe is entirely dominated by dark energy, which would imply that $H(z)$ indeed resembles a constant. In any case, a viable cosmology should at least reproduce the observed values of $H(z)$ at low redshifts, which is the regime that is studied in this work. Furthermore, at low redshifts the exact specifics of the cosmological evolution have a negligible effect on our measurements. Therefore, to calculate distances from redshifts throughout this work, we can adopt an effective $\lcdm$ background cosmology with $\Omega_{\rm m}=0.315$ and $\Omega_{\rm \Lambda}=0.685$ {\cite[]{planck2015}}, without significantly affecting our results. To calculate the distribution of apparent DM, we use the value of $H_0 = 70 \, {\rm km \, s^{-1} Mpc^{-1}}$. Throughout the paper we use the following definition for the reduced Hubble constant: $h \equiv h_{70} = H_0 / (70 \, {\rm km \, s^{-1} Mpc^{-1}})$.

Because, as mentioned above, EG gives an effective description of GR (with apparent DM as an additional component), we assume that a gravitational potential affects the pathway of photons as it does in the GR framework. This means that the distribution of apparent DM can be observed using the regular gravitational lensing formalism. In this work we test the predictions of EG specifically relating to galaxy-galaxy lensing (GGL): the coherent gravitational distortion of light from a field of background galaxies (sources) by the mass of a foreground galaxy sample (lenses) (see e.g. \citealp{fischer2000ggl,hoekstra2004,mandelbaum2006,velander2014,uitert2016}). Because the prediction of the gravitational potential in EG is currently only valid for static, spherically symmetric and isolated baryonic mass distributions, we need to select our lenses to satisfy these criteria. Furthermore, as mentioned above, the lenses should be at relatively low redshifts since cosmological evolution is not yet implemented in the theory. To find a reliable sample of relatively isolated foreground galaxies at low redshift, we select our lenses from the very complete spectroscopic Galaxy And Mass Assembly survey {\cite[GAMA,][]{driver2011gama}}. In addition, GAMA's stellar mass measurements allow us to test the prediction of EG for four galaxy sub-samples with increasing stellar mass. The background galaxies, used to measure the lensing effect, are observed by the photometric Kilo-Degree Survey {\cite[KiDS,][]{dejong2013kids}}, which was specifically designed with accurate shape measurements in mind.

In Sect. \ref{sec:gama} of this paper we explain how we select and model our lenses. In Sect. \ref{sec:measurement} we describe the lensing measurements. In Sect. \ref{sec:prediction} we introduce the EG theory and derive its prediction for the lensing signal of our galaxy sample. In Sect. \ref{sec:results} we present the measured GGL profiles and our comparison with the predictions from EG and $\lcdm$. The discussion and conclusions are described in Sect. \ref{sec:discon}.

\section{GAMA lens galaxies}
\label{sec:gama}

The prediction of the gravitational potential in EG that is tested in this work is only valid for static, spherically symmetric and isolated baryonic mass distributions (see Sect. \ref{sec:prediction}). Ideally we would like to find a sample of isolated lenses, but since galaxies are clustered we cannot use GAMA to find galaxies that are truly isolated. Instead we use the survey to construct a sample of lenses that dominate their surroundings, and a galaxy sample that allows us to estimate the small contribution arising from their nearby low-mass galaxies (i.e. satellites). The GAMA survey {\cite[]{driver2011gama}} is a spectroscopic survey with the AAOmega spectrograph mounted on the Anglo-Australian Telescope. In this study, we use the GAMA II \cite[]{liske2015gamaII} observations over three equatorial regions (G09, G12 and G15) that together span $\sim180 \deg^2$. Over these regions, the redshifts and properties of 180,960 galaxies\footnote{These are all galaxies with redshift quality $n_{\rm Q} \geq 2$. However, the recommended redshift quality of GAMA (that we use in our analysis) is $n_{\rm Q} \geq 3$.} are measured. These data have a redshift completeness of $98.5\%$ down to a Petrosian $r$-band magnitude of $m_{\rm r} = 19.8$. This is very useful to accurately determine the positional relation between galaxies, in order to find a suitable lens sample.

\subsection{Isolated galaxy selection}
\label{sec:selection}
To select foreground lens galaxies suitable for our study, we consult the 7th GAMA Galaxy Group Catalogue (G3Cv7) which is created by {\cite{robotham2011lenscat}} using a Friends-of-Friends (FoF) group finding algorithm. In this catalogue, galaxies are classified as either the Brightest Central Galaxy (BCG) or a satellite of a group, depending on their luminosity and their mutual projected and line-of-sight distances. In cases where there are no other galaxies observed within the linking lengths, the galaxy remains `non-grouped' (i.e., it is not classified as belonging to any group). Mock galaxy catalogues, which were produced using the Millennium DM simulation \cite[]{springel2005} and populated with galaxies according to the semi-analytical galaxy formation recipe `GALFORM' \cite[]{bower2006}, are used to calibrate these linking lengths and test the resulting group properties. 

\begin{figure}
	\includegraphics[width=1.0\columnwidth]{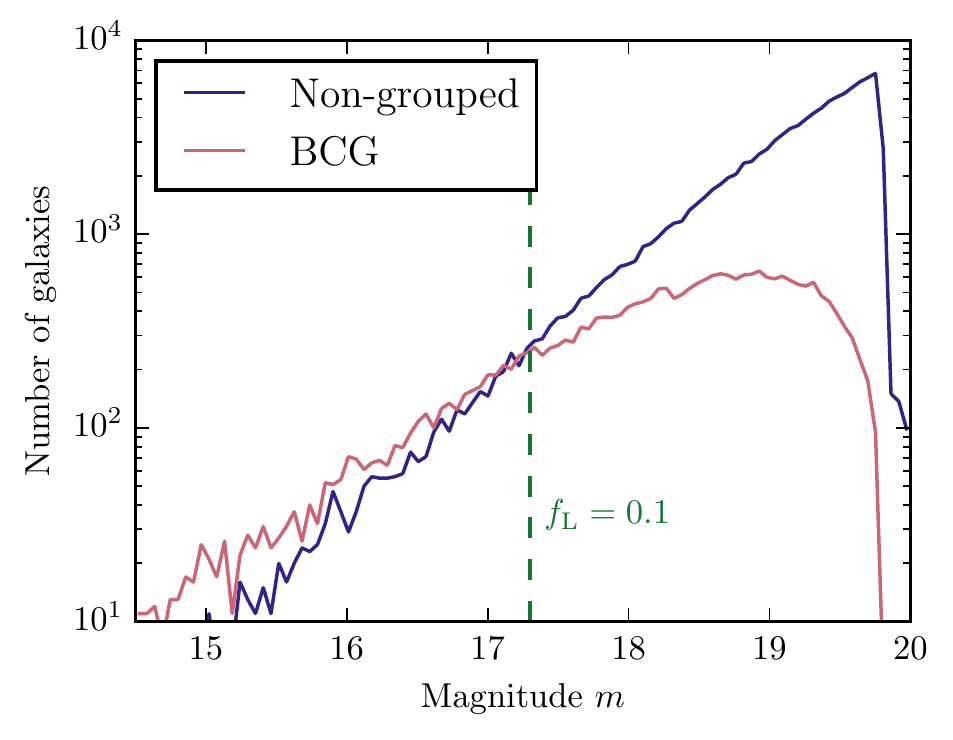}
	\caption{The magnitude distribution of non-grouped galaxies (blue) and BCGs (red). The green dashed line indicates the selection that removes galaxies which might have a satellite beyond the visible magnitude limit. These hypothetical satellites have at most a fraction $f_{\rm L} = 0.1$ of the central galaxy luminosity, corresponding to the magnitude limit: $m_{\rm r} < 17.3$. We use this `nearby' sample to obtain a reliable estimate of the satellite distribution around our centrals.}
	\label{fig:magnitude}
\end{figure}

However, since GAMA is a flux-limited survey, it does not include the satellites of the faintest observed GAMA galaxies when these are fainter than the flux limit. Many fainter galaxies are therefore classified as non-grouped, whereas they are in reality BCGs. This selection effect is illustrated in Fig. \ref{fig:magnitude}, which shows that the number of non-grouped galaxies rises towards faint magnitudes whereas the number of BCGs peaks well before. The only way to obtain a sample of `isolated' GAMA galaxies without satellites as bright as $f_L$ times their parents luminosity, would be to select only non-grouped galaxies brighter than $1/f_{\rm L}$ times the flux limit (illustrated in Fig. \ref{fig:magnitude} for $f_{\rm L}=0.1$). Unfortunately such a selection leaves too small a sample for a useful lensing measurement. Moreover, we suspect that in some cases observational limitations may have prevented the detection of satellites in this sample as well. Instead, we use this selection to obtain a reasonable estimate of the satellite distribution around the galaxies in our lens sample. Because the mass of the satellites is approximately spherically distributed around the BCG, and is sub-dominant compared to the BCG's mass, we can still model the lensing signal of this component using the EG theory. How we model the satellite distribution and its effect on the lensing signal is described in Sect. \ref{sec:satellites} and Sect. \ref{sec:extended_prediction} respectively.

Because centrals are only classified as BCGs if their satellites are detected, whereas non-grouped galaxies are likely centrals with no observed satellites, we adopt the name `centrals' for the combined sample of BCGs and non-grouped galaxies (i.e. all galaxies which are not satellites). As our lens sample, we select galaxies which dominate their surroundings in three ways: (i) they are centrals, i.e. not classified as satellites in the GAMA group catalogue; (ii) they have stellar masses below $10^{11} \hsmsun$, since we find that galaxies with higher stellar mass have significantly more satellites (see Sect. \ref{sec:satellites}); and (iii) they are not affected by massive neighbouring groups, i.e. there is no central galaxy within $3 \hsMpc$ (which is the maximum radius of our lensing measurement, see Sect. \ref{sec:measurement}). This last selection suppresses the contribution of neighbouring centrals (known as the `2-halo term' in the standard DM framework) to our lensing signal, which is visible at scales above $\sim1\hsMpc$.

Furthermore, we only select galaxies with redshift quality $n_{\rm Q} \geq 3$, in accordance with the standard recommendation by GAMA. After these four cuts (central, no neighbouring centrals, $M_* < 10^{11} \hsmsun$ and $n_{\rm Q} \geq 3$) our remaining sample of `isolated centrals' amounts to $33,613$ lenses.

\subsection{Baryonic mass distribution}
\label{sec:massdist}
Because there exists no DM component in the \cite{verlinde2016} framework of EG, the gravitational potential originates only from the baryonic mass distribution. Therefore, in order to determine the lensing signal of our galaxies as predicted by EG (see Sect. \ref{sec:prediction}), we need to know their baryonic mass distribution. In this work we consider two possible models: the point mass approximation and an extended mass profile. We expect the point mass approximation to be valid, given that (i) the bulk mass of a galaxy is enclosed within the minimum radius of our measurement ($R_{\rm min} = 30 \hskpc$), and (ii) our selection criteria ensure that our isolated centrals dominate the total mass distribution within the maximum radius of our measurement ($R_{\rm max} = 3 \hsMpc$). If these two assumptions hold, the entire mass distribution of the isolated centrals can be described by a simple point mass. This allows us to analytically calculate the lensing signal predicted by EG, based on only one observable: the galaxies' mass $M_{\rm g}$, which consists of a stellar and a cold gas component. To asses the sensitivity of our interpretation to the mass distribution, we compare the predicted lensing signal of the point mass to that of an extended mass distribution. This more realistic extended mass profile consists of four components: stars, cold gas, hot gas and satellites, which all have an extended density profile. In the following sections we review each component, and make reasonable assumptions regarding their model profiles and corresponding input parameters.

\subsubsection{Stars and cold gas}
\label{sec:stellarmass}
To determine the baryonic masses $M_{\rm g}$ of the GAMA galaxies, we use their stellar masses $M_*$ from version 19 of the stellar mass catalogue, an updated version of the catalogue created by {\cite{taylor2011mstar}}. These stellar masses are measured from observations of the Sloan Digital Sky Survey {\cite[SDSS,][]{abazajian2009sdss}} and the VISTA Kilo-Degree Infrared Galaxy survey {\cite[VIKING,][]{edge2013}}, by fitting {\cite{bruzual2003mstar}} stellar population synthesis models to the \emph{ugrizZYJHK} spectral energy distributions (constrained to the rest frame wavelength range 3,000-11,000 \AA). We correct $M_*$ for flux falling outside the automatically selected aperture using the `flux-scale' parameter, following the procedure discussed in \cite{taylor2011mstar}.

In these models, the stellar mass includes the mass locked up in stellar remnants, but not the gas recycled back into the interstellar medium. Because the mass distribution of gas in our galaxies is not measured, we can only obtain realistic estimates from literature. There are two contributions to consider: cold gas consisting of atomic hydrogen (HI), molecular hydrogen (${\rm H_2}$) and helium, and hot gas consisting of ionized hydrogen and helium. Most surveys find that the mass in cold gas is highly dependent on the galaxies' stellar mass. For low-redshift galaxies ($z<0.5$) the mass in HI (${\rm H_2}$) ranges from $20-30\%$ ($8-10\%$) of the stellar mass for galaxies with $M_*=10^{10} {\rm M_\odot}$, dropping to $5-10\%$ ($4-5\%$) for galaxies with $M_*=10^{11} {\rm M_\odot}$ \cite[]{saintonge2011,catinella2013,boselli2014,morokuma2015}. Therefore, in order to estimate the mass of the cold gas component, we consider a cold gas fraction $f_{\rm cold}$ which depends on the measured $M_*$ of our galaxies. We use the best-fit scaling relation found by \cite{boselli2014} using the Herschel Reference Survey \cite[]{boselli2010}:
\begin{equation}
\log\left(f_{\rm cold}\right) = \log\left({M_{\rm cold}}/{M_*}\right) = -0.69 \log(M_*) + 6.63 \, .
\end{equation}
In this relation, the total cold gas mass $M_{\rm cold}$ is defined as the combination of the atomic and molecular hydrogen gas, including an additional 30\% contribution of helium: $M_{\rm cold} = 1.3\left(M_{\rm HI} + M_{\rm H_2}\right)$. With a maximum measured radius of \mbox{$\sim1.5$} times the effective radius of the stellar component, the extent of the cold gas distribution is very similar to that of the stars \cite[]{pohlen2010,crocker2011,cooper2012,davis2013}. We therefore consider the stars and cold gas to form a single galactic mass distribution with:
\begin{equation}
M_{\rm g} = \left(M_* + {M_{\rm cold}}\right) = M_*(1+f_{\rm cold}) \, .
\end{equation}
For both the point mass and the extended mass profile, we use this galactic mass $M_{\rm g}$ to predict the lensing signal in the EG framework.

In the point mass approximation, the total density distribution of our galaxies consists of a point source with its mass corresponding to the galactic mass $M_{\rm g}$ of the lenses. For the extended mass profile, we use $M_{\rm g}$ as an input parameter for the density profile of the `stars and cold gas' component. Because starlight traces the mass of this component, we use the S\'{e}rsic intensity profile {\cite[]{sersic1963, sersic1968}} as a reasonable approximation of the density:
\begin{equation}
I_{\rm S}(r) \propto \rho_{\rm S}(r) = \rho_{\rm e} \, \exp\left\{ -b_{\rm n} \left[ \left(\frac{r}{r_{\rm e}} \right)^{1/n} - 1\right] \right\} \, .
\label{eq:sersic}
\end{equation}
Here $r_{\rm e}$ is the effective radius, $n$ is the S\'{e}rsic index, and $b_{\rm n}$ is defined such that $\Gamma(2n) = 2\gamma(2n, b_{\rm n})$. The S\'{e}rsic parameters were measured for $167,600$ galaxies by {\cite{kelvin2012}} on the UKIRT Infrared Deep Sky Survey Large Area Survey images from GAMA and the $ugrizYJHK$ images of SDSS DR7 (where we use the parameter values as measured in the $r$-band). Of these galaxies, $69,781$ are contained in our GAMA galaxy catalogue. Although not all galaxies used in this work (the $33,613$ isolated centrals) have S\'{e}rsic parameter measurements, we can obtain a realistic estimate of the mean S\'{e}rsic parameter values of our chosen galaxy samples. We use $r_{\rm e}$ and $n$ equal to the mean value of the galaxies for which they are measured within each sample, in order to model the density profile $\rho_{\rm S}(r)$ of each full galaxy sample. This profile is multiplied by the effective mass density $\rho_{\rm e}$, which is defined such that the mass integrated over the full $\rho_{\rm S}(r)$ is equal to the mean galactic mass $\lan M_{\rm g} \ra$ of the lens sample. The mean measured values of the galactic mass and S\'{e}rsic parameters for our galaxy samples can be found in Table \ref{tab:lenses}.

\begin{table*}
	\centering
	\caption{For each stellar mass bin, this table shows the number $N$ and mean redshift $\lan z_{\rm l}\ra$ of the galaxy sample. Next to these, it shows the corresponding measured input parameters of the ESD profiles in EG: the mean stellar mass $\lan M_* \ra$, galactic mass $\lan M_{\rm g}\ra$, effective radius $\lan r_{\rm e}\ra$, S\'{e}rsic index $\lan n\ra$, satellite fraction $\lan f_{\rm sat} \ra$ and satellite radius $\lan r_{\rm sat} \ra$ of the centrals. All masses are displayed in units of $\log_{10}(M/\hsmsun)$ and all lengths in $\hskpc$.}
	
	\begin{tabular}{lllllllll}
		\hline
		
		$M*$-bin & $N$ & $\lan z_{\rm l}\ra$ & $\lan M_*\ra$ & $\lan M_{\rm g}\ra$ & $\lan r_{\rm e}\ra$ & $\lan n\ra$ & $\lan f_{\rm sat} \ra$ & $\lan r_{\rm sat} \ra$ \\ \hline
		
		$8.5-10.5$  &  $14974$  &  $0.22$  &  $10.18$  &  $10.32$  &  $3.58$  &  $1.66$  &  $0.27$  &  $140.7$ \\
		$10.5-10.8$  &  $10500$  &  $0.29$  &  $10.67$  &  $10.74$  &  $4.64$  &  $2.25$  &  $0.25$  &  $143.9$  \\
		$10.8-10.9$  &  $4076$  &  $0.32$  &  $10.85$  &  $10.91$  &  $5.11$  &  $2.61$  &  $0.29$  &  $147.3$  \\
		$10.9-11$  &  $4063$  &  $0.33$  &  $10.95$  &  $11.00$  &  $5.56$  &  $3.04$  &  $0.32$  &  $149.0$  \\
		
		\hline
	\end{tabular}
	
	\label{tab:lenses}
\end{table*}

\subsubsection{Hot gas}
Hot gas has a more extended density distribution than stars and cold gas, and is generally modelled by the $\beta$-profile \cite[e.g.][]{cavaliere1976,mulchaey2000}:
\begin{equation}
\rho_{\rm hot}(r) = \frac{\rho_{\rm core}}{\left( 1+ \left(r/r_{\rm core} \right)^2 \right)^\frac{3\beta}{2} } \, ,
\end{equation}
which provides a fair description of X-ray observations in clusters and groups of galaxies. In this distribution $r_{\rm core}$ is the core radius of the hot gas. The outer slope is characterised by $\beta$, which for a hydrostatic isothermal sphere corresponds to the ratio of the specific energy in galaxies to that in the hot gas \cite[see e.g.][for a review]{mulchaey2000}. Observations of galaxy groups indicate $\beta\sim 0.6$ \cite[]{sun2009}. \cite{fedeli2014b} found similar results using the Overwhelmingly Large Simulations \cite[OWLS,][]{schaye2010} for the range in stellar masses that we consider here (i.e. with $M_*\sim 10^{10}-10^{11} \hsmsun$). We therefore adopt $\beta=0.6$. Moreover, \cite{fedeli2014b} estimate that the mass in hot gas is at most $3$ times that in stars. As the X-ray properties from the OWLS model of active galactic nuclei match X-ray observations well \cite[]{mccarthy2010} we adopt $M_{\rm hot}=3\langle M_*\rangle$. \cite{fedeli2014b} find that the simulations suggest a small core radius $r_{\rm core}$ (i.e. even smaller than the transition radius of the stars). This implies that $\rho_{\rm hot}(r)$ is effectively described by a single power law. Observations show a range in core radii, but typical values are tens of kpc \cite[e.g.][]{mulchaey1996} for galaxy groups. We take $r_{\rm c}=r_{\rm e}$, which is relatively small in order to give an upper limit; a larger value would reduce the contribution of hot gas, and thus move the extended mass profile closer to the point mass case. We define the amplitude $\rho_{\rm core}$ of the profile such that the mass integrated over the full $\rho_{\rm hot}(r)$ distribution is equal to the total hot gas mass $M_{\rm hot}$.

\subsubsection{Satellites}
\label{sec:satellites}
As described in \ref{sec:selection} we use our nearby ($m_{\rm r} < 17.3$) sample of centrals (BCGs and non-grouped galaxies) to find that most of the non-grouped galaxies in the GAMA catalogue might not be truly isolated, but are likely to have satellites beyond the visible magnitude limit. Fortunately, satellites are a spherically distributed, sub-dominant component of the lens, which means its (apparent) mass distribution can be described within EG. In order to assess the contribution of these satellites to our lensing signal, we first need to model their average baryonic mass distribution. We follow {\cite{uitert2016}} by modelling the density profile of satellites around the central as a double power law\footnote{Although this double power law is mathematically equivalent to the Navarro-Frenk-White profile \cite[]{navarro1995nfw} which describes virialized DM halos, it is in our case not related to any (apparent) DM distribution. It is merely an empirical fit to the measured distribution of satellite galaxies around their central galaxy.}:
\begin{equation}
\rho_{\rm sat}(r) = \frac{\rho_{\rm sat}}{(r/r_{\rm sat})(1+r/r_{\rm sat})^2}  \, ,
\label{eq:nfw_sat}
\end{equation}
where $\rho_{\rm sat}$ is the density and $r_{\rm sat}$ the scale radius of the satellite distribution. The amplitude $\rho_{\rm sat}$ is chosen such that the mass integrated over the full profile is equal to the mean total mass in satellites $\lan M^{\rm sat}_* \ra$ measured around our nearby sample of centrals. By binning these centrals according to their stellar mass $M^{\rm cen}_*$ we find that, for centrals within $10^{9} < M^{\rm cen}_* < 10^{11} \hsmsun$, the total mass in satellites can be approximated by a fraction $f_{\rm sat} = \lan{M^{\rm sat}_*}\ra/\lan{M^{\rm cen}_*}\ra \sim0.2-0.3$. However, for centrals with masses above $10^{11} \hsmsun$ the satellite mass fraction rapidly rises to $f_{\rm sat}\sim1$ and higher. For this reason, we choose to limit our lens sample to galaxies below $10^{11} \hsmsun$. As the value of the scale radius $r_{\rm sat}$, we pick the half-mass radius (the radius which contains half of the total mass) of the satellites around the nearby centrals. The mean measured mass fraction $\lan f_{\rm sat} \ra$ and half-mass radius $\lan r_{\rm sat} \ra$ of satellites around centrals in our four $M_*$-bins can be found in Table \ref{tab:lenses}.

\section{Lensing Measurement}
\label{sec:measurement}

According to GR, the gravitational potential of a mass distribution leaves an imprint on the path of travelling photons. As discussed in Sect. \ref{sec:introduction}, EG gives an effective description of GR (where the excess gravity from apparent DM detailed in {\citealp{verlinde2016}} is an additional component). We therefore work under the assumption that a gravitational potential (including that of the predicted apparent DM distribution) has the exact same effect on light rays as in GR. Thus, by measuring the coherent distortion of the images of faraway galaxies (sources), we can reconstruct the projected (apparent) mass distribution (lens) between the background sources and the observer. In the case of GGL, a large sample of foreground galaxies acts as the gravitational lens (for a general introduction, see e.g. {\citealp{bartelmann2001lensing, schneider2006lensing}}). Because the distortion of the source images is only $\sim 1 \%$ of their intrinsic shape, the tangential shear $\gamma_t$ (which is the source ellipticity tangential to the line connecting the source and the centre of the lens) is averaged for many sources within circular annuli around the lens centre. This measurement provides us with the average shear $\lan\gamma_{\rm t}\ra(R)$ as a function of projected radial distance $R$ from the lens centres. In GR, this quantity is related to the Excess Surface Density (ESD) profile $\Delta\Sigma(R)$. Using our earlier assumption, we can also use the same methodology to obtain the ESD of the apparent DM in the EG framework. The ESD is defined as the average surface mass density $\lan\Sigma\ra(<R)$ within $R$, minus the surface density $\Sigma(R)$ at that radius:
\begin{equation}
	\Delta\Sigma (R) = \lan\Sigma\ra(<R) - \Sigma (R) = \lan \gamma_{\rm t} \ra (R) \, \Sigma_{\rm crit} \, .
	\label{eq:deltasigma}
\end{equation}
Here $\Sigma_{\rm crit}$ is the critical surface mass density at the redshift of the lens:
\begin{equation}
	\Sigma_{\rm crit} = \frac{c^2}{4\pi G} \frac{D(z_{\rm s})}{D(z_{\rm l}) \, D(z_{\rm l}, z_{\rm s})} \, ,
	\label{eq:sigmacrit}
\end{equation}
a geometrical factor that is inversely proportional to the strength of the lensing effect. In this equation $D(z_{\rm l})$ and $D(z_{\rm s})$ are the angular diameter distances to the lens and source respectively, and $D(z_{\rm l}, z_{\rm s})$ is the distance between the lens and the source.

For a more extensive discussion of the GGL method and the role of the KiDS and GAMA surveys therein, we refer the reader to previous KiDS-GAMA lensing papers:  \cite{sifon2015, uitert2016, brouwer2016} and especially Sect. 3 of {\cite{viola2015}}.

\subsection{KiDS source galaxies}
\label{sec:kids}

The background sources used in our GGL measurements are observed by KiDS {\cite[]{dejong2013kids}}. The KiDS photometric survey uses the OmegaCAM instrument {\cite[]{kuijken2011omegacam}} on the VLT Survey Telescope {\cite[]{capaccioli2011vlt}} which was designed to provide a round and uniform point spread function (PSF) over a square degree field of view, specifically with weak lensing measurements in mind. Of the currently available $454 \deg^2$ area from the `KiDS-450' data release {\cite[]{hildebrandt2016}} we use the $\sim 180 \deg^2$ area that overlaps with the equatorial GAMA fields {\cite[]{driver2011gama}}. After masking bright stars and image defects, $79 \%$ of our original survey overlap remains \cite[]{dejong2015}.

The photometric redshifts of the background sources are determined from $ugri$ photometry as described in \mbox{\cite{kuijken2015}} and \cite{hildebrandt2016}. Due to the bias inherent in measuring the source redshift probability distribution $p(z_{\rm s})$ of each individual source (as was done in the previous KiDS-GAMA studies), we instead employ the source redshift number distribution $n(z_{\rm s})$ of the full population of sources. The individual $p(z_{\rm s})$ is still measured, but only to find the `best' redshift $z_{\rm B}$ at the $p(z_{\rm s})$-peak of each source. Following \cite{hildebrandt2016} we limit the source sample to: $z_{\rm B} < 0.9$. We also use $z_{\rm B}$ in order to select sources which lie sufficiently far behind the lens: $z_{\rm B} > z_{\rm l}+0.2$. The $n(z_{\rm s})$ is estimated from a spectroscopic redshift sample, which is re-weighted to resemble the photometric properties of the appropriate KiDS galaxies for different lens redshifts (for details, see Sect. 3 of \citealp{hildebrandt2016} and \citealp{uitert2016}). We use the $n(z)$ distribution behind the lens for the calculation of the critical surface density from Eq. (\ref{eq:sigmacrit}):
\begin{equation}
	\Sigma_{\rm crit}^{-1} = \frac{4\pi G}{c^2} D(z_{\rm l}) \intop_{z_{\rm l}+0.2}^{\infty} \frac{D(z_{\rm l}, z_{\rm s})}{D(z_{\rm s})} n(z_{\rm l},z_{\rm s}) \, {\rm d} z_{\rm s} \, ,
\end{equation}
By assuming that the intrinsic ellipticities of the sources are randomly oriented, $\lan \gamma_{\rm t}\ra$ from Eq. (\ref{eq:deltasigma}) can be approximated by the average tangential ellipticity $\lan \epsilon_{\rm t}\ra$ given by:
\begin{equation}
	\epsilon_{\rm t} = -\epsilon_1 \cos(2\phi) - \epsilon_2 \sin(2\phi) \, ,
\end{equation}
where $\epsilon_1$ and $\epsilon_2$ are the measured source ellipticity components, and $\phi$ is the angle of the source relative to the lens centre (both with respect to the equatorial coordinate system). The measurement of the source ellipticities is performed on the $r$-band data, which is observed under superior observing conditions compared to the other bands {\cite[]{dejong2015,kuijken2015}}. The images are reduced by the {\scshape theli} pipeline ({\citealp{erben2013theli}} as described in {\citealp{hildebrandt2016}}). The sources are detected from the reduced images using the {\scshape SExtractor} algorithm {\cite[]{bertin1996sextractor}}, whereafter the ellipticities of the source galaxies are measured using the improved self-calibrating \emph{lens}fit code {\cite[]{miller2007lensfit, miller2013lensfit, fenechconti2016}}. Each shape is assigned a weight $w_{\rm s}$ that reflects the reliability of the ellipticity measurement. We incorporate this \emph{lens}fit weight and the lensing efficiency $\Sigma_{\rm crit}^{-1}$ into the total weight:
\begin{equation}
	W_{\rm{ls}} = w_{\rm s} \Sigma_{\rm crit}^{-2} \, ,
	\label{eq:weights}
\end{equation}
which is applied to each lens-source pair. This factor down-weights the contribution of sources that have less reliable shape measurements, and of lenses with a redshift closer to that of the sources (which makes them less sensitive to the lensing effect).

Inside each radial bin $R$, the weights and tangential ellipticities of all lens-source pairs are combined according to Eq. (\ref{eq:deltasigma}) to arrive at the ESD profile:
\begin{equation}
	\Delta\Sigma (R) = \frac{1}{1+K} \frac{\sum_{ls} W_{ls} \epsilon_{\rm t} \Sigma_{{\rm crit},l} }{ \sum_{ls}{W_{ls}} } \, .
	\label{eq:ESDmeasured}
\end{equation}
In this equation, $K$ is the average correction of the multiplicative bias $m$ on the \emph{lens}fit shear estimates. The values of $m$ are determined using image simulations \cite[]{fenechconti2016} for 8 tomographic redshift slices between $0.1 \leq z_{\rm B} < 0.9$ (Dvornik et al., in prep). The average correction is computed for the lens-source pairs in each respective redshift slice as follows:
\begin{equation}
K=\frac{\sum_{ls} W_{ls} m_{s}}{\sum_{ls} W_{ls}} \, ,
\end{equation}
where the mean value of $K$ over the entire source redshift range is $-0.014$.

We also correct the ESD for systematic effects that arise from the residual shape correlations due to PSF anisotropy. This results in non-vanishing contributions to the ESD signal on large scales and at the survey edges, because the averaging is not done over all azimuthal angles. This spurious signal can be determined by measuring the lensing signal around random points. We use $\sim18$ million locations from the GAMA random catalogue, and find that the resulting signal is small (below $10\%$ for scales up to $\sim 1 \hsMpc$). We subtract the lensing signal around random locations from all measured ESD profiles.
	
Following previous KiDS-GAMA lensing papers, we measure the ESD profile for 10 logarithmically spaced radial bins within $0.02 < R < 2 \, h_{100}^{-1} {\rm Mpc}$, where our estimates of the signal and uncertainty are thoroughly tested\footnote{\cite{viola2015} used the following definition of the reduced Hubble constant: $h \equiv h_{100} = H_0 / (100 \, {\rm km \, s^{-1} Mpc^{-1}})$}. However, since we work with the $h \equiv h_{70}$ definition, we use the approximately equivalent $0.03 < R < 3 \, h_{70}^{-1} {\rm Mpc}$ as our radial distance range. The errors on the ESD values are given by the diagonal of the analytical covariance matrix. Section 3.4 of {\cite{viola2015}} includes the computation of the analytical covariance matrix and shows that, up to a projected radius of $R=2 \, h_{100}^{-1} {\rm Mpc}$, the square root of the diagonal is in agreement with the error estimate from bootstrapping.

\section{Lensing Signal Prediction}
\label{sec:prediction}

According to \cite{verlinde2016}, the gravitational potential $\Phi(r)$ caused by the enclosed baryonic mass distribution $M_{\rm b}(r)$ exceeds that of GR on galactic and larger scales. In addition to the normal GR contribution of $M_{\rm b}(r)$ to $\Phi(r)$, there exists an extra gravitational effect. This excess gravity arises due to a volume law contribution to the entropy that is associated with the positive dark energy in our universe. In a universe without matter the total entropy of the dark energy would be maximal, as it would be non-locally distributed over all available space. In our universe, on the other hand, any baryonic mass distribution $M_{\rm b}(r)$ reduces the entropy content of the universe. This removal of entropy due to matter produces an elastic response of the underlying microscopic system, which can be observed on galactic and larger scales as an additional gravitational force. Although this excess gravity does not originate from an actual DM contribution, it can be effectively described by an \emph{apparent} DM distribution $M_{\rm D}(r)$.

\subsection{The apparent dark matter formula}
\cite{verlinde2016} determines the amount of apparent DM by estimating the elastic energy associated with the entropy displacement caused by $M_{\rm b}(r)$. This leads to the following relation\footnote{Although \cite{verlinde2016} derives his relations for an arbitrary number of dimensions $d$, for the derivation in this paper we restrict ourselves to four spacetime dimensions.}:
\begin{equation}
	\int_0^r \varepsilon_{\rm D}^2 (r') A(r')dr' = V_{M_{\rm b}} (r)  \, ,
	\label{eq:fundamentalrelation}
\end{equation}
where we integrate over a sphere with radius $r$ and area $A(r) = 4 \pi r^2$. The strain $\varepsilon_{\rm D}(r)$ caused by the entropy displacement is given by:
\begin{equation}
	\varepsilon_{\rm D} (r) = \frac{8\pi G}{c H_0} \frac{M_{\rm D}(r)}{A(r)} \, ,
	\label{eq:strain}
\end{equation}
where $c$ is the speed of light, $G$ the gravitational constant, and $H_0$ the present-day Hubble constant (which we choose to be $H_0= 70 \, {\rm km \, s^{-1} Mpc^{-1}}$). Furthermore, $V_{M_{\rm b}}(r)$ is the volume that would contain the amount of entropy that is removed by a mass $M_{\rm b}$ inside a sphere of radius $r$, if that volume were filled with the average entropy density of the universe:
\begin{equation} 
	V_{M_{\rm b}}(r) = \frac{8\pi G}{c H_0} \frac{M_{\rm b} (r) \, r}{3} \, .
	\label{eq:removedvolume}
\end{equation}
Now inserting the relations (\ref{eq:strain}) and (\ref{eq:removedvolume}) into (\ref{eq:fundamentalrelation}) yields:
\begin{equation}
	\int_0^r \frac{G M_{\rm D}^2 (r')}{r'^2} dr' = M_{\rm b}(r) r \frac{c H_0}{6} \, .
\end{equation}
Finally, by taking the derivative with respect to $r$ on both sides of the equation, one arrives at the following relation:
\begin{equation}
	M_{\rm D}^2 (r) = \frac{  cH_0 r^2}{6G} \frac{d\left( M_{\rm b}(r) r \right)}{dr} \, .
	\label{eq:veg_mdm}
\end{equation}
This is the apparent DM formula from \cite{verlinde2016}, which translates a baryonic mass distribution into an apparent DM distribution. This apparent DM only plays a role in the regime where the elastic response of the entropy of dark energy $S_{\rm DE}$ takes place: where $V(r) > V_{M_{\rm b}}(r)$, i.e. $S_{\rm DE}\propto V(r)$ is large compared to the entropy that is removed by $M_{\rm b}(r)$ within our radius $r$. By substituting Eq. (\ref{eq:removedvolume}) into this condition, we find that this is the case when:
\begin{equation}
	r > \sqrt{\frac{2 G}{c H_0} {M_{\rm b} (r)}} \, .
	\label{eq:regime}
\end{equation}
For a lower limit on this radius for our sample, we can consider a point source with a mass of $M=10^{10} \hsmsun$, close to the average mass $\lan M_{\rm g} \ra$ of galaxies in our lowest stellar mass bin. In this simple case, the regime starts when $r>2\hskpc$. This shows that our observations (which start at $30\hskpc$) are well within the EG regime. 

However, it is important to keep in mind that this equation does not represent a new fundamental law of gravity, but is merely a macroscopic approximation used to describe an underlying microscopic phenomenon. Therefore, this equation is only valid under the specific set of circumstances that have been assumed for its derivation. In this case, the system considered was a static, spherically symmetric and isolated baryonic mass distribution. With these limitations in mind, we have selected our galaxy sample to meet these criteria as closely as possible (see Sect. \ref{sec:selection}).

Finally we note that, in order to test the EG predictions with gravitational lensing, we need to make some assumptions about the used cosmology (as discussed in Sect. \ref{sec:introduction}). These concern the geometric factors in the lensing equation (Eq. \ref{eq:sigmacrit}), and the evolution of the Hubble constant (which enters in Eq. (\ref{eq:veg_mdm}) for the apparent DM). We assume that, if EG is to be a viable theory, it should predict an expansion history that agrees with the current supernova data \cite[]{riess1996,kessler2009,betoule2014}, specifically over the redshift range that is relevant for our lensing measurements ($0.2 < z_{\rm s} < 0.9$). If this is the case, the angular diameter distance-redshift relation is similar to what is used in $\lcdm$. We therefore adopt a $\lcdm$ background cosmology with $\Omega_{\rm m}=0.315$ and $\Omega_{\rm \Lambda}=0.685$, based on the {\cite{planck2015}} measurements. Regarding $H_0$ in Eq. (\ref{eq:veg_mdm}), we note that a Hubble parameter that changes with redshift is not yet implemented in the EG theory. However, for the lens redshifts considered in this work ($\lan z_{\rm l}\ra \sim 0.2$) the difference resulting from using $H_0$ or $H(z_{\rm l})$ to compute the lensing signal prediction is $\sim5\%$. This means that, considering the statistical uncertainties in our measurements ($\gtrsim 40\%$, see e.g. Fig. \ref{fig:extended_model}), our choice to use $H_0 = 70 \, {\rm km \, s^{-1} Mpc^{-1}}$ instead of an evolving $H(z_{\rm l})$ has no significant effect on the results of this work.

From Eq. (\ref{eq:veg_mdm}) we now need to determine the ESD profile of the apparent DM distribution, in order to compare the predictions from EG to our measured GGL profiles. The next steps toward this $\Delta\Sigma_{\rm EG}(R)$ depend on our assumptions regarding the baryonic mass distribution of our lenses. We compute the lensing signal in EG for two models (which are discussed in Sect. \ref{sec:massdist}): the point mass approximation and the more realistic extended mass distribution.

\subsection{Point mass approximation}
\label{sec:pointsource}

In this work we measure the ESD profiles of galaxies at projected radial distances $R>30 \hskpc$. If we assume that, beyond this distance, the galaxy is almost entirely enclosed within the radius $r$, we can approximate the enclosed baryonic mass as a constant: $M_{\rm b}(r) = M_{\rm b}$. Re-writing Eq. (\ref{eq:veg_mdm}) accordingly yields:
\begin{equation}
	M_{\rm D}(r)=\sqrt{\frac{cH_{0}}{6 \, G}}\, r \sqrt{M_{\rm b}} \equiv C_{\rm D} \, r \sqrt{M_{\rm b}} \, ,
\end{equation}
where $C_{\rm D}$ is a constant factor determined by $c$, $G$ and $H_0$. In order to calculate the resulting $\Delta\Sigma_{\rm D}(R)$ we first need to determine the spherical density distribution $\rho_{\rm D}(r)$. Under the assumption of spherical symmetry, we can use:
\begin{equation}
	\rho_{\rm D}(r) = \frac{1}{4\pi r^{2}} \frac{dM_{\rm D}(r)}{dr} = \frac{C_{\rm D}\sqrt{M_{\rm b}}}{4\pi r^{2}} \, .
	\label{eq:Md_to_rhod}
\end{equation}
We calculate the corresponding surface density $\Sigma_{\rm D}(R)$ as a function of projected distance $R$ in the cylindrical coordinate system ($R$, $\phi$, $z$), where $z$ is the distance along the line-of-sight and $r^2=R^2+z^2$, such that:
\begin{equation}
	\Sigma_{\rm D}(R)=\intop_{-\infty}^{\infty}\rho_{\rm D}(R,z) \, {\rm d}z \, .
	\label{eq:projected}
\end{equation}
Substituting $\rho_{\rm D}(R,z)$ provides the surface density of the apparent DM distribution associated with our point mass:
\begin{equation}
	\Sigma_{\rm D}(R) = \frac{C_{\rm D}\sqrt{M_{\rm b}}}{4\pi} \, 2 \intop_{0}^{\infty} \frac{{\rm d} z}{R^{2}+z^{2}} = \frac{C_{\rm D}\sqrt{M_{\rm b}}}{4R} \, .
	\label{eq:veg_sigma}
\end{equation}
We can now use Eq. (\ref{eq:deltasigma}) to find the ESD:
\begin{multline}
	\Delta\Sigma(R) = {\lan\Sigma\ra}(<R)-\Sigma(R) = \\ \frac{2\pi\intop_{0}^{R} \Sigma(R') R' \, {\rm d}R'}{\pi R^{2}} - \Sigma(R) \, .
	\label{eq:Sigma_to_DSigma}
\end{multline}
In the case of our point mass:
\begin{equation}
	\Delta\Sigma_{\rm D}(R) = \frac{C_{\rm D}\sqrt{M_{\rm b}}}{2R} - \frac{C_{\rm D}\sqrt{M_{\rm b}}}{4R}  = \frac{C_{\rm D}\sqrt{M_{\rm b}}}{4R} \, ,
\end{equation}
which happens to be equal to $\Sigma_{\rm D}(R)$ from Eq. (\ref{eq:veg_sigma})\footnote{Note that the ESD of the apparent DM distribution: $\Delta\Sigma_{\rm D}(R) \propto \sqrt{H_0 M_{\rm b}}/R \propto \sqrt{h}$, is explicitly dependent on the Hubble constant, which means that an incorrect measured value of $H_0$ would affect our conclusions.}.

Apart from the extra contribution from the apparent DM predicted by EG, we also need to add the standard GR contribution from baryonic matter to the ESD. Under the assumption that the galaxy is a point mass we know that $\Sigma_{\rm b}(R)=0$ for $R>0$, and that the integral over $\Sigma_{\rm b}(R)$ must give the total mass $M_{\rm b}$ of the galaxy. Substituting this into Eq. (\ref{eq:Sigma_to_DSigma}) gives us:
\begin{equation}
\Delta\Sigma_{\rm b}(R) = \frac{M_{\rm b}}{\pi R^2} \, .
\label{eq:pointmass}
\end{equation}
Ultimately, the total ESD predicted by EG in the point mass approximation is:
\begin{equation}
\Delta\Sigma_{\rm EG}(R) = \Delta\Sigma_{\rm b}(R) + \Delta\Sigma_{\rm D}(R) \, ,
\label{eq:veg_esd}
\end{equation}
where the contributions are the ESDs of a point source with mass $M_{\rm g}$ of our galaxies, both in GR and EG.

\subsection{Extended mass distribution}
\label{sec:extended_prediction}

The above derivation only holds under the assumption that our galaxies can be considered point masses. To test whether this is justified, we wish to compare the point mass prediction to a more realistic lens model. This model includes the extended density profile for stars, cold gas, hot gas and satellites as described in Sect. \ref{sec:massdist}. To determine the ESD profile of the extended galaxy model as predicted by EG, we cannot perform an analytical calculation as we did for the point mass approximation. Instead we need to calculate the apparent DM distribution $M^{\rm ext}_{\rm D}(r)$ and the resulting $\Delta\Sigma^{\rm ext}_{\rm D}(R)$ numerically for the sum of all baryonic components. We start out with the total spherical density distribution $\rho^{\rm ext}_{\rm b}(r)$ of all components:
\begin{equation}
	\rho^{\rm ext}_{\rm b}(r) = \rho^{\rm S}_{\rm b}(r) + \rho^{\rm hot}_{\rm b}(r) + \rho^{\rm sat}_{\rm b}(r) \, ,
\end{equation}
where the respective contributions are: the S\'{e}rsic model for stars and cold gas, the $\beta$-profile for hot gas, and the double power law for satellites. We numerically convert this to the enclosed mass distribution:
\begin{equation}
	M^{\rm ext}_{\rm b}(r) = 4 \pi \intop_0^r \rho^{\rm ext}_{\rm b}(r') r'^2 \, {\rm d}r' \, .
\end{equation}
We rewrite Eq. (\ref{eq:veg_mdm}) in order to translate $M^{\rm ext}_{\rm b}(r)$ to its corresponding distribution of apparent DM in EG:
\begin{equation}
	M^{\rm ext}_{\rm D}(r)= C_{\rm D} r \, \sqrt{ \frac{d M^{\rm ext}_{\rm b}(r) \, r }{d r} } \, ,
\end{equation}
which is numerically converted into the apparent DM density distribution $\rho^{\rm ext}_{\rm D}(r)$ by substituting $M^{\rm ext}_{\rm D}(r)$ into Eq. (\ref{eq:Md_to_rhod}).

The projected surface density $\Sigma^{\rm ext}_{\rm D}(R)$ from Eq. (\ref{eq:projected}) is calculated by computing the value of $\rho^{\rm ext}_{\rm D}(R,z)$ in cylindrical coordinates for $10^3$ values of $z$ and integrating over them. The last step towards computing the ESD profile is the subtraction of $\Sigma^{\rm ext}_{\rm D}(R)$ from the average surface density within $R$, as in Eq. (\ref{eq:Sigma_to_DSigma}), where ${\lan\Sigma^{\rm ext}_{\rm D}\ra}(<R)$ is calculated by performing the cumulative sum over $2\pi R \, \Sigma^{\rm ext}_{\rm D}(R)$ and dividing the result by its cumulative area. In addition to the lensing signal from apparent DM, we need to include the baryonic ESD profile. We numerically compute $\Delta\Sigma^{\rm ext}_{\rm b}(R)$ from $\rho^{\rm ext}_{\rm b}(r)$ in the same way as we computed $\Delta\Sigma^{\rm ext}_{\rm D}(R)$ from $\rho^{\rm ext}_{\rm D}(r)$. This makes the total ESD predicted by EG for the extended mass distribution:
\begin{equation}
\Delta\Sigma^{\rm ext}_{\rm EG}(R) = \Delta\Sigma^{\rm ext}_{\rm b}(R) + \Delta\Sigma^{\rm ext}_{\rm D}(R) \, .
\label{eq:ext_esd}
\end{equation}

When considering the resulting ESD profiles of the extended density models, we must keep in mind that they only represent reasonable estimates which contain uncertainties for two different reasons:
\begin{enumerate}
	
	\item The extended baryonic density distribution of each component is approximated using reasonable assumptions on the used model profiles and their corresponding input parameters. These assumptions are based on observations of the galaxies in our sample and of other galaxies, and also on simulations. Although we try to find suitable input parameters corresponding to the measured stellar mass of our galaxy samples, we cannot be certain that our modelled density distributions are completely correct.
	
	\item We cannot model the extended density distribution for each individual GAMA galaxy, but have to assume one average profile per lens sample (based on the average stellar mass $\lan M_* \ra$ of that sample). Translating the extended baryonic mass model to the lensing profile of its corresponding apparent DM distribution (as explained above) is a highly non-linear operation. Therefore, we cannot be certain that the calculated lensing profile of an average density distribution is exactly the same as the lensing profile of all individual galaxies combined, although these would only differ greatly in the unlikely case that there is a large spread in the input parameters of the extended mass profiles within each stellar mass sub-sample.
	
\end{enumerate}
For these two reasons we cannot use the average profile as a reliable model for the apparent DM lensing signal of our galaxy samples. In the point mass approximation, we do have the measured input parameter (the stellar mass) for each individual galaxy, and we can compute the apparent DM lensing profile for each individual galaxy. However, this approach can only be used when the contribution from hot gas and satellites is small. We therefore compare our estimate of the apparent DM lensing profile of the extended mass distribution to that of the point masses, to assess the error margins in our EG prediction.

\begin{figure*}
	\includegraphics[width=1.0\textwidth]{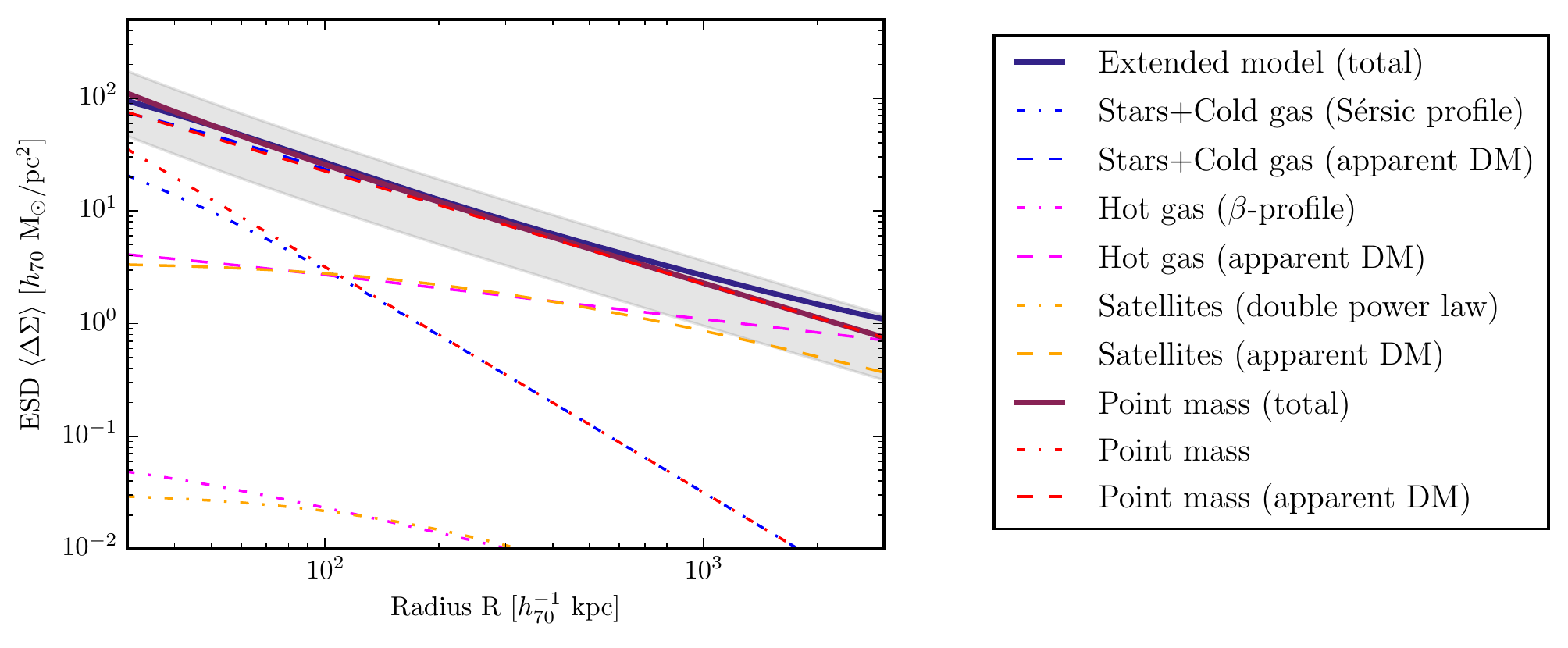}
	\caption{The ESD profile predicted by EG for isolated centrals, both in the case of the point mass approximation (dark red, solid) and the extended galaxy model (dark blue, solid). The former consists of a point source with the mass of the stars and cold gas component (red), with the lensing signal evaluated for both the baryonic mass (dash-dotted) and the apparent DM (dashed). The latter consists of a stars and cold gas component modelled by a S\'{e}rsic profile (blue), a hot gas component modelled by a $\beta$-profile (magenta), and a satellite distribution modelled by a double power law (orange), all with the lensing signal evaluated for both the baryonic mass (dash-dotted) and the apparent DM (dashed). Note that the total ESD of the extended mass distribution is not equal to the sum of its components, due to the non-linear conversion from baryonic mass to apparent DM. All profiles are shown for our highest mass bin ($10^{10.9} < M_* < 10^{11} \hsmsun$), but the difference between the two models is similar for the other galaxy sub-samples. The difference between the ESD predictions of the two models is comparable to the median $1 \sigma$ uncertainty on our lensing measurements (illustrated by the grey band).}
	\label{fig:extended_model}
\end{figure*}

The total ESD profile predicted for the extended density distribution, and that of each component\footnote{Note that, due to the non-linear nature of the calculation of the apparent DM distribution, the total ESD profile of the extended mass distribution is not the sum of the components shown in \mbox{Fig. \ref{fig:extended_model}}.}, is shown in Fig. \ref{fig:extended_model}. We only show the profiles for the galaxies in our highest stellar mass bin: $10^{10.9} < M_* < 10^{11} \hsmsun$, but since the relations between the mass in hot gas, satellites and their galaxies are approximately linear, the profiles look similar for the other sub-samples. At larger scales, we find that the point mass approximation predicts a lower ESD than the extended mass profile. However, the difference between the $\Delta\Sigma(R)$ predictions of these two models is comparable to the median $1 \sigma$ uncertainty on the ESD of our sample (which is illustrated by the gray band in Fig. \ref{fig:extended_model}). We conclude that, given the current statistical uncertainties in the lensing measurements, the point mass approximation is adequate for isolated centrals within the used radial distance range ($0.03 < R < 3\hsMpc$).

\section{Results}
\label{sec:results}

\begin{figure*}
	\centering
	\includegraphics[width=0.8\textwidth]{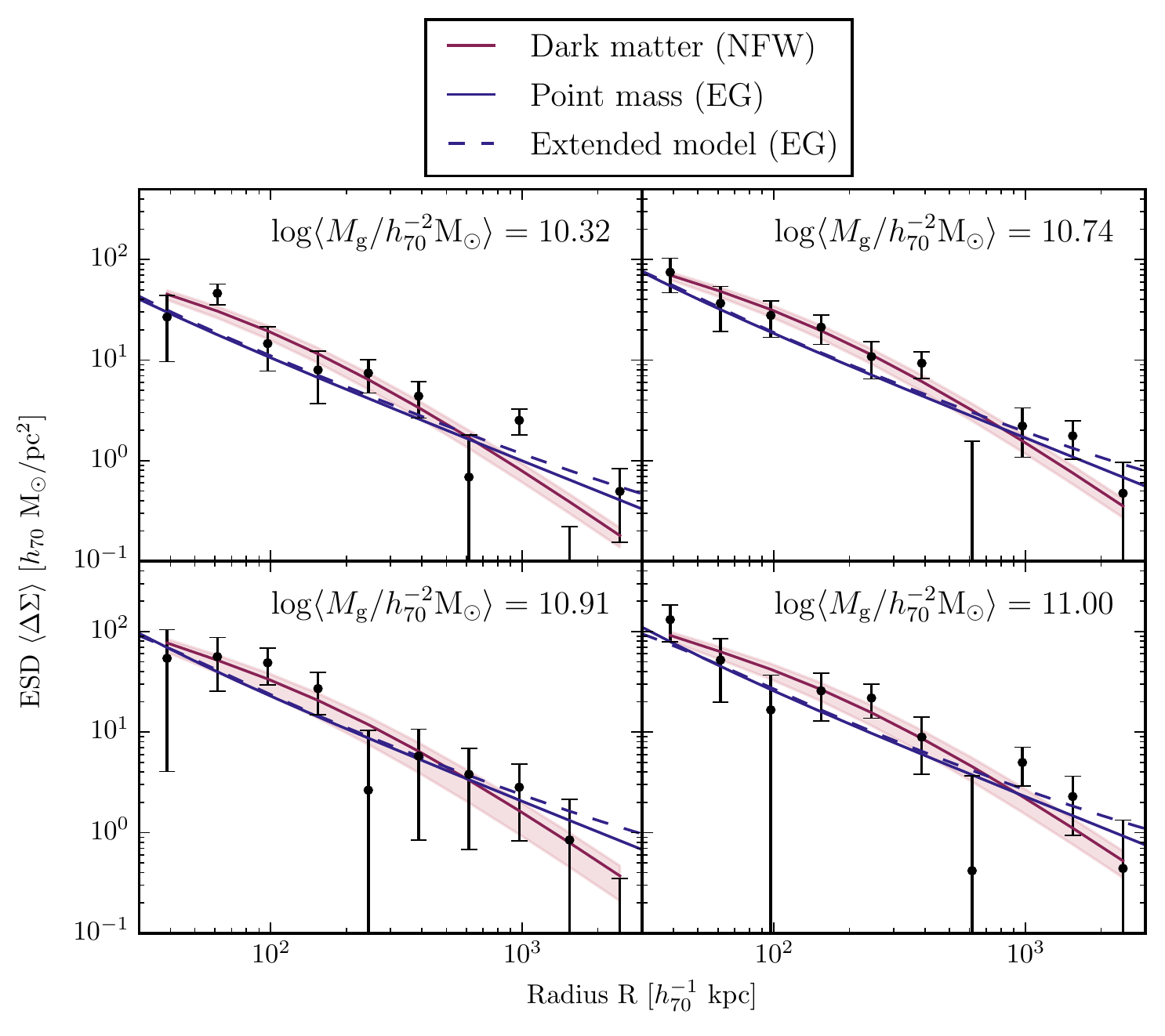}
	\caption{The measured ESD profiles of isolated centrals with $1 \sigma$ error bars (black), compared to those predicted by EG in the point mass approximation (blue) and for the extended mass profile (blue, dashed). Note that \emph{none of these predictions are fitted to the data}: they follow directly from the EG theory by substitution of the baryonic masses $M_{\rm g}$ of the galaxies in each sample (and, in the case of the extended mass profile, reasonable assumptions for the other baryonic mass distributions). The mean measured galaxy mass is indicated at the top of each panel. For comparison we show the ESD profile of a simple NFW profile as predicted by GR (red), with the DM halo mass $M_{\rm h}$ fitted as a free parameter in each stellar mass bin.}
	\label{fig:results}
\end{figure*}

We measure the ESD profiles (following Sect. \ref{sec:measurement}) of our sample of isolated centrals, divided into four sub-samples of increasing stellar mass. The boundaries of the $M_*$-bins: $\log(M_*/\hsmsun) = [8.5, \, 10.5, \, 10.8, \, 10.9, \, 11.0]$, are chosen to maintain an approximately equal signal-to-noise in each bin. Figure \ref{fig:results} shows the measured ESD profiles (with $1\sigma$ error bars) of galaxies in the four $M_*$-bins. Together with these measurements we show the ESD profile predicted by EG, under the assumption that our isolated centrals can be considered point masses at scales within $0.03 < R < 3\hsMpc$. The masses $M_{\rm g}$ of the galaxies in each bin serve as input in Eq. (\ref{eq:veg_esd}), which provides the ESD profiles predicted by EG for each individual galaxy. The mean baryonic masses of the galaxies in each $M_*$-bin can be found in Table \ref{tab:lenses}. The ESDs of the galaxies in each sample are averaged to obtain the total $\Delta\Sigma_{\rm EG}(R)$. It is important to note that the shown EG profiles do \emph{not contain any free parameters}: both their slope and amplitudes are fixed by the prediction from the EG theory (as stated in Eq. \ref{eq:veg_mdm}) and the measured masses $M_{\rm g}$ of the galaxies in each $M_*$-bin. Although this is only a first attempt at testing the EG theory using lensing data, we can perform a very simple comparison of this prediction with both the lensing observations and the prediction from the standard $\lcdm$ model.

\begin{table}
\centering
\caption{For each stellar mass bin, this table shows the median values (including 16th and 84th percentile error margins) of the halo mass $M_{\rm h}$ obtained by the NFW fit, and the `best' amplitude $A_{\rm B}$ that minimizes the $\chi^2$ if the EG profile were multiplied by it (for the point mass and extended mass profile). The halo masses are displayed in units of $\log_{10}(M/\hsmsun)$.}
	
\begin{tabular}{llll}
	\hline
	$M*$-bin     & $M_{\rm h}$ & $A_{\rm B}$ & $A^{\rm ext}_{\rm B}$ \\ \hline
	
	$8.5-10.5$  &  $12.15^{+0.10}_{-0.11}$  &  $1.36^{+0.21}_{-0.21}$  &  $1.21^{+0.19}_{-0.19}$  \\
	$10.5-10.8$  &  $12.45^{+0.10}_{-0.11}$  &  $1.32^{+0.19}_{-0.19}$  &  $1.20^{+0.18}_{-0.18}$  \\
	$10.8-10.9$  &  $12.43^{+0.17}_{-0.22}$  &  $1.07^{+0.27}_{-0.27}$  &  $0.94^{+0.25}_{-0.25}$  \\
	$10.9-11$  &  $12.62^{+0.13}_{-0.16}$  &  $1.33^{+0.25}_{-0.26}$  &  $1.20^{+0.23}_{-0.24}$  \\
		
	\hline
\end{tabular}
	
\label{tab:output}
\end{table}

\subsection{Model comparison}

In standard GGL studies performed within the $\lcdm$ framework, the measured ESD profile is modelled by two components: the baryonic mass of the galaxy and its surrounding DM halo. The baryonic component is often modelled as a point source with the mean baryonic mass of the galaxy sample, whereas the DM halo component usually contains several free parameters, such as the mass and concentration of the halo, which are evaluated by fitting a model to the observed ESD profiles. Motivated by N-body simulations, the DM halo is most frequently modelled by the Navarro-Frenk-White density profile {\cite[NFW,][]{navarro1995nfw}}, very similar to the double power law in Eq. (\ref{eq:nfw_sat}). This profile has two free parameters: the halo mass $M_{\rm h}$, which gives the amplitude, and the scale radius $r_{\rm s}$, which determines where the slope changes. Following previous GAMA-KiDS lensing papers (see e.g. {\citealp{sifon2015,viola2015,uitert2016,brouwer2016}}) we define $M_{\rm h}$ as $M_{200}$: the virial mass contained within $r_{200}$, and we define the scale radius in terms of the concentration: $c\equiv r_{200}/r_{\rm s}$. In these definitions, $r_{200}$ is the radius that encloses a density of $200$ times $\rho_{\rm m}(z)$, the average matter density of the universe. Using the {\cite{duffy2008massrelation}} mass-concentration relation, we can define $c$ in terms of $M_{\rm h}$. We translate the resulting density profile, which depends exclusively on the DM halo mass, into the projected ESD distribution following the analytical description of {\cite{wright2000model}}. We combine this NFW model with a point mass that models the baryonic galaxy component (as in Eq. \ref{eq:pointmass}). Because our lens selection minimizes the contribution from neighbouring centrals (see Sect. \ref{sec:selection}), we do not need to add a component that fits the 2-halo term. We fit the NFW model to our measured ESD profiles using the {\scshape emcee} sampler {\cite[]{foreman2013emcee}} with 100 walkers performing 1000 steps. The model returns the median posterior values of $M_{\rm h}$ (including 16th and 84th percentile error margins) displayed in Table \ref{tab:output}. The best-fit ESD profile of the NFW model (including 16th and 84th percentile bands) is shown in Fig. \ref{fig:results}.

For both the $\Delta\Sigma_{\rm EG}$ predicted by EG (in the point mass approximation) and the simple NFW fit $\Delta\Sigma_{\rm NFW}$, we can compare the $\Delta\Sigma_{\rm mod}$ of the model with the observed $\Delta\Sigma_{\rm obs}$ by calculating the $\chi^2$ value:
\begin{multline}
	\chi^2 = (\Delta\Sigma_{\rm obs} - \Delta\Sigma_{\rm mod})^\intercal \cdot C^{-1}(\Delta\Sigma_{\rm obs} - \Delta\Sigma_{\rm mod}) \, ,
	\label{eq:chi2}
\end{multline}
where $C^{-1}$ is the inverse of the analytical covariance matrix (see Sect. \ref{sec:measurement}). 
From this quantity we can calculate the reduced $\chi^2$ statistic\footnote{While the reduced $\chi^2$ statistic is shown to be a suboptimal goodness-of-fit estimator (see e.g. \citealp{andrae2010}) it is a widely used criterion, and we therefore discuss it here for completeness.}: $\chi_{\rm red}^2 = {\chi^2}/{N_{\rm DOF}}$. It depends on the number of degrees of freedom (DOF) of the model: $N_{\rm DOF} = N_{\rm data} - N_{\rm param}$, where $N_{\rm data}$ is the number of data-points in the measurement and $N_{\rm param}$ is the number of free parameters. Due to our choice of 10 $R$-bins and 4 $M_*$-bins, we use $4\times10 = 40$ data-points. In the case of EG there are no free parameters, which means $N^{\rm EG}_{\rm DOF} = 40$. Our simple NFW model has one free parameter $M_{\rm h}$ for each $M_*$-bin, resulting in $N^{\rm NFW}_{\rm DOF} = 40-4 = 36$. The resulting total $\chi_{\rm red}^2$ over the four $M_*$-bins is $44.82/40=1.121$ for EG, and $33.58/36=0.933$ for the NFW fit. In other words, both the NFW and EG prediction agree quite well with the measured ESD profile, where the NFW fit has a slightly better $\chi_{\rm red}^2$ value. Since the NFW profile is an empirical description of the surface density of virialized systems, the apparent correspondence of both the NFW fit and the EG prediction with the observed ESD essentially reflects that the predicted EG profile roughly follows that of virialized systems.

A more appropriate way to compare the two models, however, is in the Bayesian framework. We use a very simple Bayesian approach by computing the Bayesian Information Criterion {\cite[$BIC$,][]{schwarz1978}}. This criterion, which is based on the maximum likelihood $\mathcal{L}_{\rm max}$ of the data given a model, penalizes model complexity more strongly than the $\chi_{\rm red}^2$. This model comparison method is closely related to other information criteria such as the Akaike Information Criterion \cite[$AIK$,][]{akaike1973} which have become popular because they only require the likelihood at its maximum value, rather than in the whole parameter space, to perform a model comparison \cite[see e.g.][]{liddle2007}. This approximation only holds when the posterior distribution is Gaussian and the data points are independent. Calculating the $BIC$, which is defined as:
\begin{equation}
BIC = -2 \ln(\mathcal{L}_{\rm max}) + N_{\rm param} \ln(N_{\rm data}) \, , 
\end{equation}
allows us to consider the relative evidence of two competing models, where the one with the lowest $BIC$ is preferred. The difference $\Delta BIC$ gives the significance of evidence against the higher $BIC$, ranging from ``0 - 2: Not worth more than a bare mention" to ``>10: Very strong" \cite[]{kass1995}. In the Gaussian case, the likelihood can be rewritten as: $-2 \ln(\mathcal{L}_{\rm max}) = \chi^2$. Using this method, we find that $BIC_{\rm EG} = 44.82$ and $BIC_{\rm NFW} = 48.33$. This shows that, when the number of free parameters is taken into account, the EG model performs at least as well as the NFW fit. However, in order to really distinguish between these two models, we need to reduce the uncertainties in our measurement, in our lens modelling, and in the assumptions related to EG theory and halo model.

In order to further assess the quality of the EG prediction across the $M_*$-range, we determine the `best' amplitude $A_{\rm B}$ and index $n_{\rm B}$: the factors that minimize the $\chi^2$ statistic when we fit:
\begin{equation}
	\Delta\Sigma_{\rm EG}(A_{\rm B}, n_{\rm B}, R) = A_{\rm B}\frac{C_{\rm D}\sqrt{M_{\rm b}}}{4} \, \left(\frac{R}{\hskpc}\right)^{-n_{\rm B}} \, ,
\end{equation}
We find that the slope of the EG prediction is very close to the observed slope of the ESD profiles, with a mean value of $\lan n_{\rm B} \ra = 1.01^{+0.02}_{-0.03}$. In order to obtain better constraints on $A_{\rm B}$, we set $n_{\rm B}=1$. The values of $A_{\rm B}$ (with $1 \sigma$ errors) for the point mass are shown in Table \ref{tab:output}. We find the amplitude of the point mass prediction to be consistently lower than the measurement. This is expected since the point mass approximation only takes the mass contribution of the central galaxy into account, and not that of extended components like hot gas and satellites (described in Sect. \ref{sec:massdist}). However, the ESD of the extended profile (which is shown in Fig. \ref{fig:results} for comparison) does not completely solve this problem. When we determine the best amplitude for the extended mass distribution by scaling its predicted ESD, we find that the values of $A^{\rm ext}_{\rm B}$ are still larger than 1, but less so than for the point mass (at a level of $\sim1\sigma$, see Table \ref{tab:output}). Nevertheless, the comparison of the extended ESD with the measured lensing profile yields a slightly higher reduced $\chi^2$: $45.50/40=1.138$. However, accurately predicting the baryonic and apparent DM contribution of the extended density distribution is challenging (see Sect. \ref{sec:extended_prediction}). Therefore, the extended ESD profile can primarily be used as an indication of the uncertainty in the lens model.

\section{Conclusion}
\label{sec:discon}

Using the $\sim180 \deg^2$ overlap of the KiDS and GAMA surveys, we present the first test of the theory of emergent gravity proposed in \cite{verlinde2016} using weak gravitational lensing. In this theory, there exists an additional component to the gravitational potential of a baryonic mass, which can be described as an \emph{apparent} DM distribution. Because the prediction of the apparent DM profile as a function of baryonic mass is currently only valid for static, spherically symmetric and isolated mass distributions, we select $33,613$ central galaxies that dominate their surrounding mass distribution, and have no other centrals within the maximum radius of our measurement ($R_{\rm max} = 3 \hsMpc$). We model the baryonic matter distribution of our galaxies using two different assumptions for their mass distribution: the point mass approximation and the extended mass profile. In the point mass approximation we assume that the bulk of the galaxy's mass resides within the minimum radius of our measurement ($R_{\rm min} = 30 \hskpc$), and model the lens as a point source with the mass of the stars and cold gas of the galaxy. For the extended distribution, we not only model the stars and cold gas component as a S\'{e}rsic profile, but also try to make reasonable estimates of the extended hot gas and satellite distributions. We compute the lensing profiles of both models and find that, given the current statistical uncertainties in our lensing measurements, both models give an adequate description of isolated centrals. In this regime (where the mass distribution can be approximated by point mass) the lensing profile of apparent DM in EG is the same as that of the excess gravity in MOND\footnote{After this paper was accepted for publication, it was pointed out to us that \cite{milgrom2013} showed that galaxy-galaxy lensing measurements from the Canada-France-Hawaii Telescope Legacy Survey \cite[performed by][]{brimioulle2013} are consistent with predictions from relativistic extensions of MOND up to a radius of $140 \, h_{72}^{-1} {\rm kpc}$ (note added in proof).}, for the specific value $a_0 = c H_0 / 6$.

When computing the observed and predicted ESD profiles, we need to make several assumptions concerning the EG theory. The first is that, because EG gives an effective description of GR in empty space, the effect of the gravitational potential on light rays remains unchanged. This allows us to use the regular gravitational lensing formalism to measure the ESD profiles of apparent DM in EG. Our second assumption involves the used background cosmology. Because EG is only developed for present-day de Sitter space, we need to assume that the evolution of cosmological distances is approximately equal to that in $\lcdm$, with the cosmological parameters as measured by \cite{planck2015}. For the relatively low redshifts used in this work ($0.2<z_{\rm s}<0.9$), this is a reasonable assumption. The third assumption is the value of $H_0$ that we use to calculate the apparent DM profile from the baryonic mass distribution. In an evolving universe, the Hubble parameter $H(z)$ is expected to change as a function of the redshift $z$. This evolution is not yet implemented in EG. Instead it uses the approximation that we live in a dark energy dominated universe, where $H(z)$ resembles a constant. We follow \cite{verlinde2016} by assuming a constant value, in our case: $H_0 = 70 \, {\rm km \, s^{-1} Mpc^{-1}}$, which is reasonable at a mean lens redshift of $\lan z_{\rm l} \ra \sim 0.2$. However, in order to obtain a more accurate prediction for the cosmology and the lensing signal in the EG framework, all these issues need to be resolved in the future.

Using the mentioned assumptions, we measure the ESD profiles of isolated centrals in four different stellar mass bins, and compare these with the ESD profiles predicted by EG. They exhibit a remarkable agreement, especially considering that the predictions contain \emph{no free parameters}: both the slope and the amplitudes within the four $M_*$-bins are completely fixed by the EG theory and the measured baryonic masses $M_{\rm g}$ of the galaxies. In order to perform a very simple comparison with $\lcdm$, we fit the ESD profile of a simple NFW distribution (combined with a baryonic point mass) to the measured lensing profiles. This NFW model contains one free parameter, the halo mass $M_{\rm h}$, for each stellar mass bin. We compare the reduced $\chi^2$ of the NFW fit (which has 4 free parameters in total) with that of the prediction from EG (which has no free parameters). Although the NFW fit has fewer degrees of freedom (which slightly penalizes $\chi^2_{\rm red}$) the reduced $\chi^2$ of this model is slightly lower than that of EG, where $\chi^2_{\rm red, NFW}=0.933$ and $\chi^2_{\rm red, EG}=1.121$ in the point mass approximation. For both theories, the value of the reduced $\chi^2$ is well within reasonable limits, especially considering the very simple implementation of both models. The fact that our observed density profiles resemble both NFW profiles and the prediction from EG, suggests that this theory predicts a phenomenology very similar to a virialized DM halo. Using the Bayesian Information Criterion, we find that $BIC_{\rm EG} = 44.82$ and $BIC_{\rm NFW} = 48.33$. These $BIC$ values imply that, taking the number of data points and free parameters into account, the EG prediction describes our data at least as well as the NFW fit. However, a thorough and fair comparison between $\lcdm$ and EG would require a more sophisticated implementation of both theories, and a full Bayesian analysis which properly takes the free parameters and priors of the NFW model into account. Nonetheless, given that the model uncertainties are also addressed, future data should be able to distinguish between the two theories.

We propose that this analysis should not only be carried out for this specific case, but on multiple scales and using a variety of different probes. From comparing the predictions of EG to observations of isolated centrals, we need to expand our studies to the scales of larger galaxy groups, clusters, and eventually to cosmological scales: the cosmic web, BAO's and the CMB power spectrum. Furthermore, there are various challenges for EG, especially concerning observations of dynamical systems such as the Bullet Cluster {\cite[]{randall2008}} where the dominant mass component appears to be separate from the dominant baryonic component. There is also ongoing research to assess whether there exists an increasing mass-to-light ratio for galaxies of later type {\cite[]{martinsson2013}}, which might challenge EG if confirmed. We conclude that, although this first result is quite remarkable, it is only a first step. There is still a long way to go, for both the theoretical groundwork and observational tests, before EG can be considered a fully developed and solidly tested theory. In this first GGL study, however, EG appears to be a good parameter-free description of our observations.

\section*{Acknowledgements}

M. Brouwer and M. Visser would like to thank Erik Verlinde for helpful clarifications and discussions regarding his emergent gravity theory. We also thank the anonymous referee for the useful comments, that helped to improve this paper.

The work of M. Visser was supported by the ERC Advanced Grant 268088-EMERGRAV, and is part of the Delta ITP consortium, a program of the NWO. M. Bilicki, H. Hoekstra and C. Sif\'on acknowledge  support  from  the  European  Research Council  under  FP7  grant  number  279396. K. Kuijken is supported by the Alexander von Humboldt Foundation. M. Bilicki acknowledges support from the Netherlands Organisation for Scientific Research (NWO) through grant number 614.001.103. H. Hildebrandt is supported by an Emmy Noether grant (No. Hi 1495/2-1) of the Deutsche Forschungsgemeinschaft. R. Nakajima acknowledges support from the German Federal Ministry for Economic Affairs and Energy (BMWi) provided via DLR under project no. 50QE1103. Dominik Klaes is supported by the Deutsche Forschungsgemeinschaft in the framework of the TR33 `The Dark Universe'.

This research is based  on  data  products  from  observations  made  with  ESO Telescopes at the La Silla Paranal Observatory under programme IDs 177.A-3016, 177.A-3017 and 177.A-3018, and on data products produced by Target OmegaCEN, INAF-OACN, INAF-OAPD and  the  KiDS  production  team,  on  behalf  of  the  KiDS  consortium. OmegaCEN and the KiDS production team acknowledge support by NOVA and NWO-M grants. Members of INAF-OAPD and INAF-OACN also acknowledge the support from the Department of Physics \& Astronomy of the University of Padova, and of the Department of Physics of Univ. Federico II (Naples).

GAMA is a joint European-Australasian project based around a spectroscopic campaign using the Anglo-Australian Telescope. The GAMA input catalogue is based on data taken from the Sloan Digital Sky Survey and the UKIRT Infrared Deep Sky Survey. Complementary imaging  of  the  GAMA  regions  is  being  obtained  by  a  number  of  independent survey programs including GALEX MIS, VST KiDS, VISTA VIKING, WISE, Herschel-ATLAS, GMRT and ASKAP providing UV to radio coverage. GAMA is funded by the STFC (UK), the ARC (Australia), the AAO, and the participating institutions. The GAMA website is \url{www.gama-survey.org}.

This work has made use of {\scshape python} (\url{www.python.org}), including the packages {\scshape numpy} (\url{www.numpy.org}), {\scshape scipy} (\url{www.scipy.org}) and {\scshape ipython} \cite[]{perez2007ipython}. Plots have been produced with {\scshape matplotlib} \cite[]{hunter2007matplotlib}.

\emph{Author contributions:} All authors contributed to the development and writing of this paper. The authorship list is given in three groups: the lead authors (M. Brouwer \& M. Visser), followed by two alphabetical groups. The first alphabetical group includes those who are key contributors to both the scientific analysis and the data products. The second group covers those who have either made a significant contribution to the data products, or to the scientific analysis.




\bibliographystyle{mnras}
\bibliography{biblio}



\bsp	
\label{lastpage}
\end{document}